\documentclass[copyright]{eptcs}

\providecommand{\event}{ICLP 2021} 

\usepackage{xspace}
\usepackage{latexsym}
\usepackage{amssymb}   
\usepackage{ifthen}
\newboolean{commentsaon}         
\setboolean{commentsaon}{true}
  \setboolean{commentsaon}{false}  

\newboolean{commentson} 
\setboolean{commentson}{true}

\newcommand{\comment}[1]
{\ifthenelse{\boolean{commentson}\AND\boolean{commentsaon}}
   {{\par\noindent\mbox{}{\footnotesize\blue[ *** #1 ]\par}\noindent\par}}{}}

\newcommand{\commenta}[1]
{\ifthenelse{\boolean{commentsaon}}
   {{\par\noindent\mbox{}{\small\color[rgb]{0, .5, 0}[ *** #1 ]\par}\noindent\par}}{}}

\usepackage[yyyymmdd]{datetime}

\usepackage{color}
\newcommand\blue     {\color{blue}}

\usepackage{bm}  

\usepackage{underscore}  
\usepackage{breakurl}             

\newtheorem{theorem}{Theorem}
\newtheorem{corollary}[theorem]{Corollary}
\newtheorem{lemma}[theorem]{Lemma}
\newtheorem{proposition}[theorem]{Proposition}
\newtheorem{example}[theorem]{Example}
\newtheorem{definition}[theorem]{Definition}
\newcommand*{\seq}[2][n]  {{#2_{1}, \allowbreak \ldots, \allowbreak #2_{#1}}}

\renewcommand*{\S}{{\ensuremath{\cal S}}\xspace}

\newcommand*{\NN}{{\ensuremath{\mathbb{N}}}\xspace}

\newcommand*{\Var}{{\ensuremath{\it\hspace{-1.5pt} Var}}\xspace}
\newcommand*{\VarOut}{{\ensuremath{\it VarOut}}\xspace}
\newcommand*{\VarIn}{{\ensuremath{\it VarIn}}\xspace}
\newcommand*{\Ran}{{\ensuremath{\it\hspace{-1pt} Ran}}\xspace}
\newcommand*{\Dom}{{\ensuremath{\it Dom}}\xspace}

\newcommand*{\myunderscore}{\mbox{\tt\symbol{95}}}

\newcommand*{\mybackslash}{\ensuremath{\mbox{\tt \symbol{92}}}\xspace}
\newcommand*{\myeq}{\mathop{\doteq}}
\newcommand*{\MYEQ}{\mathop{\stackrel{..}=}}

\newcommand*{\nqueens}{{\sc nqueens}\xspace}

\title{A Note on Occur-Check}
\author{W{\l}odzimierz Drabent
  \institute{Institute of Computer Science,  Polish Academy of Sciences}
  \institute{Department of Computer and Information Science,
          Link\"oping University, Sweden}
  \email{drabent\,{\it at}\/\,ipipan\,{\it dot}\/\,waw\,{\it dot}\/\,pl%
     {\ifthenelse{\boolean{commentson}\AND\boolean{commentsaon}}
       {\blue\footnotesize\quad  [with private comments, Version 1.18]}{}%
     }%
  }  
}  

\newcommand*\titlerunning{Occur-Check}
\newcommand*\authorrunning{W. Drabent}

\begin{document}
\maketitle

\begin{abstract}

  Most known results on avoiding the occur-check are based on the notion
  of ``not subject to occur-check'' (NSTO).
  It means that unification is performed only on such pairs of atoms
  for which the occur-check never succeeds in any run of a nondeterministic
  unification algorithm. 
  Here we show that this requirement is too strong.  
  We show how to weaken it, and present some related sufficient
  conditions under which the occur-check may be safely omitted.
  We show examples for which the proposed approach provides more general
  results than the approaches based on well-moded and nicely moded programs
  (this includes cases to which the latter approaches are inapplicable).
\\[.5ex]
\end{abstract}

\section{Introduction}

\noindent
The programming language Prolog implements SLD-resolution employing an unsound
implementation of unification without the occur-check.  This usually creates
no problems in practice.  Programmers know that they do not need to care
about it, unless they deal with something unusual like checking a difference
list for emptiness.%
\footnote{%
  In the important Prolog textbook by Sterling and Shapiro
  \cite{Sterling-Shapiro-shorter}, the occur-check is mentioned
  (in the context of actual programs) only when discussing difference lists
  (p.\,298, p.\,300, on p.\,299 an error due to unsound Prolog unification
   is explained).  
  The textbook of Bratko \cite{Bratko.4ed} mentions the occur-check only
  once,
  when comparing matching in Prolog with unification in logic.
}
Surprisingly, such attitude of programmers is often not justified by theory.
The known criteria for occur-check freeness are applicable to restricted
classes of cases.
There seems to exist no further substantial work on avoiding the occur-check
after that of Chadha and Plaisted \cite{ChadhaP94}, 
Apt and Pellegrini \cite{AptP94-occur-check}, reported in
\cite{apt-prolog}, and the generalization in \cite{AptL95.delays}
to other selection rules than that of Prolog.

Even for LD-resolution %
(SLD-resolution with the
Prolog selection rule) the proposed methods are 
inapplicable to some important cases.  To deal with simple examples of
programs employing difference lists, the methods of well-moded and  nicely
moded programs 
had to be refined  \cite{AptP94-occur-check}
in a rather sophisticated way.
(The refinement is not presented in the textbook \cite{apt-prolog}, 
one may suppose that it was considered too complicated.)

Here we are interested in sufficient conditions for safe execution 
of definite clause programs without the occur check.
Approaches based on semantic analysis, like abstract interpretation, are left
outside of the scope of this paper.

The existing approaches are based on the notion of NSTO
(not subject to occur-check) \cite{DBLP:conf/slp/DeransartFT91}.  It means
that unification is performed only on such pairs of atoms
for which the occur-check never succeeds in {\em any} run of a nondeterministic
unification algorithm.

It turns out that unification without the occur-check works
correctly also for some cases which are not NSTO.  In this paper we propose
a generalization of NSTO.  We show that it is sufficient that the
occur-check does not succeed in {\em one} run of the unification algorithm
for a given input (instead of {\em all} the runs).
We discuss some related sufficient conditions %
for safely avoiding the occur-check.
They are applicable to some examples to which the former approaches are
inapplicable.  For other examples, a wider class of initial queries is dealt
with, or/and applying the proposed approach seems simpler than the former
ones. 
We additionally present a sufficient condition, based on NSTO, for safely
avoiding the
occur-check under arbitrary selection rule (and provide a detailed proof of
its correctness).

\subsection*{Preliminaries}

We use the terminology, notation and many definitions  from \cite{apt-prolog}
(and reintroduce here only some of them).
The terminology is based on that of logic; in particular ``atom'' means an
atomic formula.

By an {\em expression} we mean a term, an atom, or a tuple of terms (or atoms).
An equation is a construct $s\myeq t$, where $s,t$ are expressions.
Given sequences of terms (or atoms) $s=\seq s$ and $t=\seq t$, 
the set $\{s_1\myeq t_1,\ldots,s_n\myeq t_n\}$ will be sometimes denoted by
$s\MYEQ t$.
A syntactic object (expression, equation, substitution, etc) is 
{\em linear} when no variable occurs in it more than once.
As in Prolog, each occurrence of \myunderscore\ in a syntactic object
will stand for a distinct variable.
Otherwise variable names begin with upper case letters.
$\Var(t)$ denotes the set of variables occurring in a syntactic object $t$.
We say that $s$ and $t$ are {\em variable disjoint} if 
$\Var(s)\cap\Var(t)=\emptyset$.

For a substitution $\theta = \{X_1/t_1,\ldots,X_n/t_n\}$, 
we define
$\Dom(\theta)=\{\seq X\}$,\, $\Ran(\theta)=\Var(\seq t)$, 
$\theta|S=\{X/t\in\theta \mid X\in S\,\}$
 (for a set $S$ of variables), and
$\theta|u= \theta| \Var(u)$
 (for an expression $u$).

We employ
the Martelli-Montanari unification algorithm (MMA) 
(cf.\ \cite{apt-prolog}).
It unifies a set of equations, 
by iteratively applying one of the actions below
to an equation from the current set, until no action is applicable.  
The equation is chosen nondeterministically.
\[
\begin{array}{l@{\ }l@{\ \ } l l}
(1) & f(\seq s)\myeq f (\seq t)
    &  \longrightarrow &
    \parbox[t]{.37\textwidth}{replace by equations
      $s_1\myeq t_1,\ldots,s_n\myeq t_n$
    }
\\
(2) & f(\seq s)\myeq g(\seq[m] t)
      \makebox{ where $f\neq g$}
    &  \longrightarrow &  \mbox{halt with failure}
\\
(3) & X\mathop{\doteq} X  &  \longrightarrow
        & \mbox{delete the equation}
\\
(4) & t\mathop{\doteq}X 
  \mbox{ where $t$ is not a variable}
    &  \longrightarrow & \mbox{replace by }X\myeq t
\\
(5) &
    X\mathop{\doteq}t \mbox{ where }
    \begin{tabular}[t]{@{}l@{}}
      $X\not\in\Var(t)$ and  $X$ occurs elsewhere
    \end{tabular}
    &  \longrightarrow
    &
    \parbox[t]{.36\textwidth}{
      apply substitution $\{X/t\}$ to all other equations
    }
\\
(6) & X\mathop{\doteq}t \mbox{ where $X\in\Var(t)$ }
     \makebox[0pt][l]{and $X\neq t$}
      &  \longrightarrow & \mbox{halt with failure}
\end{array}
\]
By a {\em run} of MMA for an input $E$ we mean a maximal sequence
$E_0,\ldots,E_n$ of equation sets such that $E=E_0$ and
 $E_{i}$ is obtained from $E_{i-1}$ by 
one of the actions of the algorithm ($i=1,\ldots,n$).
  See \cite{apt-prolog} for the properties of MMA, in particular how the
 obtained mgu is represented by a final equation set.

An equation set $E$ is said to be {\em NSTO}\/
if action (6) is not performed in any execution of MMA starting with $E$.
We often say ``unification of $s$ and $t$ is NSTO'' instead of 
``$\{s\myeq t\}$ is NSTO''.
In such case while unifying $s,t$
the occur check never succeeds, and thus can be skipped.

We need to generalize some definitions from \cite{apt-prolog}, 
in order not to be limited to LD-resolution.
We will say that {\em unification of $A$ and $H$ is available} in an
SLD-derivation (or SLD-tree)
for a program $P$,
if $A$ is the selected atom in a query of the derivation (tree), and $H$
is a standardized 
apart head of a clause from $P$,
such that $A$ and $H$ have the same predicate symbol.
(A more formal phrasing is {``equation set $\{A{\doteq}H\}$ is available''}.)
If all the unification{}s available in an SLD-derivation (SLD-tree) are NSTO
 then the derivation (tree) is  {\em occur-check free}.
We say that a program $P$ with a query $Q$ is {\em occur-check free}
if, under a given selection rule, the SLD-tree for $P$ with $Q$ 
is occur-check free.
\commenta{Do we need occur-check free derivations (trees)?

Later on we generalize this notion; whenever the two versions should be
distinguished the current one will be called {\em strongly occur-check free}.

If an SLD-tree for a program $P$ with a query $Q$  
is occur-check free, then $P$ with $Q$ under the given selection rule
is {\em strongly occur-check free} (occur-check free in the terminology
of \cite{apt-prolog}).  We will often skip ``strongly'', 
unless it is necessary to distinguish this notion from its generalization, 
introduced later on.
}

We refer a few times to results of \cite{AptP94-occur-check} reported in 
\cite{apt-prolog}; in such cases only a reference to \cite{apt-prolog}
may be given.

Similarly to \cite{AptP94-occur-check}, we will employ modes.
This means dividing the argument positions of predicates into two groups,
by assigning a function $m_p\colon \{1,\ldots,n\}\to\{+,-\}$ (called a
{\em mode}) to each
predicate $p$ of the considered program (where $n$ is the arity of $p$).
A program with a mode for each predicate is called a {\em moded program}, 
and the collection of modes is called {\em moding}.
We follow the usual terminology, 
argument positions with $+$ assigned are called input, and those with $-$ are
called output.
We usually specify $m_p$ by writing $p(m_p(1),\ldots,m_p(n))$.
E.g.\ $p(+,-)$ states that the first argument of $p$ is input and the second
one is output.
Note that moding does not need to correspond to any intuitive notion of data
flow. 
It is to be chosen so that the moded program satisfies the conditions of interest.

We will write $p(s;t)$ to represent an atom $p(\seq t)$ and to state that $s$ is
the sequence of terms in its input positions, and $t$ the sequence of terms
in its output positions.
 An atom
 $p(s;t)$ is {\em input-output disjoint} if $\Var(s)\cap\Var(t)=\emptyset$.
Let us define $\VarIn(p(s;t))=\Var(s)$,  $\VarOut(p(s;t))=\Var(t)$.
The input (resp.\ output) positions of a query $Q$ are the input (output)
positions of the atoms of $Q$.
A query (or an atom) $Q$ is
  {\em input linear} (resp.\ {\em output linear})
if the sequence of the terms occurring in the
input (output) positions of $Q$ is linear.  
  We will refer to the following results.

\begin{lemma}
\label{lemma.nsto}
  \begin{enumerate}
  \item 
    \label{lemma.nsto.7.14}
    Consider atoms $A$ and $H$.  If they are variable disjoint, one of them is
    input-output disjoint, one of them is input linear, and the other is output
    linear then $\{A\myeq H\}$ is NSTO  \cite[Lemma 7.14]{apt-prolog}.

  \item
    \label{lemma.nsto.7.5part}
    Let $s$ and $t$ be sequences of terms, 
  such that the lengths of $s$ and $t$ are the same.
  If\, $\Var(s)\cap\Var(t)=\emptyset$ and $s$ 
    (or $t$) is linear
    then $s\MYEQ t$ is NSTO
    (a special case of \cite[Lemma 7.5]{apt-prolog}).

  \end{enumerate}
\end{lemma}

  Obviously, $p(s)\myeq p(t)$ is NSTO iff $s\MYEQ t$ is NSTO.
Based on Lemma \ref{lemma.nsto}.\ref{lemma.nsto.7.14}, Apt and Pellegrini
\cite{AptP94-occur-check} introduced two sufficient conditions for
occur-check freeness. 
One ({\em well-moded} programs
\DeclareRobustCommand{\myfragment}{Def.\linebreak[3]\,7.8}%
\cite[\myfragment]{apt-prolog})
implies that the input positions of the atoms selected in LD-trees are ground.
The other one ({\em nicely-moded} programs
\cite[Def.\,7.19]{apt-prolog}) implies that the selected atoms are
output-linear.

\section{NSTO and arbitrary selection rules}
\subsection{Sufficient condition}
 
Here we propose a syntactic condition for occur-check freeness under
arbitrary selection rules. 
We assume that the programs dealt with are moded.

\begin{definition}%
  Let  $Q=\seq A$  be a query.
We define a relation $\to_Q$ on $\{\seq A\}$.  Let
$A_i\to_Q A_j$ when a variable occurs in an output position of $A_i$ and an
input position of~$A_j$.

  Query $Q$ is {\bf tidy} if it is output linear and $\to_Q$ is acyclic
  ($A_i\not\to_Q^+A_i$ for $i=1,\ldots,n$).

  Clause $H\gets Q$ is {\bf tidy} if $Q$ is tidy, and
  
\qquad  $H$ is input linear,  

\qquad
  no variable from an input position of $H$ occurs in an output position of $Q$.

\end{definition}

Note that each atom in a tidy query is input-output disjoint.
Also, if a query $Q$ is tidy then any permutation of $Q$ is tidy too.
A linear query is tidy under any moding.

Tidy programs
are a generalization of nicely moded programs
\cite{ChadhaP94}, \cite{AptP94-occur-check}
\cite{apt-prolog}
(the first reference uses different terminology).
A nicely moded query $\seq A$ is tidy (as
$A_i\to_Q A_j$ implies $i<j$), and a nicely moded clause with an input linear
head is tidy.
A tidy program can be converted into a nicely moded one
by reordering the atoms in the clause bodies.  
The generalization proposed here may be understood as minor.
However it still seems to be worth presenting, as the proof of the crucial
lemma in  \cite{AptL95.delays} seems unavailable.

\comment{$\!$$\!$%
It may be supposed that Krzysztof Apt himself might have some doubts about that
 generalization.$\!\!$%
}

This is a basic property of tidy clauses and queries:

\begin{lemma}
\label{lemma.tidy}
  Let $Q$ be a tidy query, and $C$ a tidy clause
  variable disjoint from $Q$.
  An SLD-resolvent $Q'$ of $Q$ and $C$ is tidy.
\end{lemma}
Due to a space limit, the proof must be excluded from this paper and
will be made available elsewhere.  We only mention here two main lemmas
involved in the proof.

\begin{lemma}
\label{lemma.3variables}
Let $\theta$ be a substitution and $X,V,V'$ variables.  Assume that
$X\in\Var(V\theta)$ and $X\in\Var(V'\theta)$.  Then $X=V=V'$, or 
$X,V,V'\in\Var(\theta)$ and moreover $X\in\Ran(\theta)$.
\end{lemma}

The next lemma employs the following notion:
A substitution
$\theta$ is {\bf linear for} a set of variables
$\S$ if for any two distinct variables 
$X,Y\in {\S}$ the pair $X\theta,Y\theta$ is linear.
Note that if a $\theta=\{X_1/t_1,\ldots,X_n/t_n\}$ is linear for $\Dom(\theta)$
then $\theta$ is linear 
in the sense of \cite[Def.\,A.3]{AptP94-occur-check}
 (i.e.\ $\seq t$ is linear).

\begin{lemma}
\label{lemma.likeA5}
Consider two variable disjoint expressions $s,t$ where $t$ is linear.  
  Let $ \S$ be a set of variables such that $\S\cap\Var(t)=\emptyset $.
If $s,t$ are unifiable then there exists a relevant and idempotent mgu $\theta$
of $s,t$ such that
\\\hspace*{5em}
  \begin{tabular}[t]{l}
    $\theta|s$ is linear for $\S$%
    \qquad and     \qquad
    $\Ran(\theta|s) \subseteq \Var(t)$.
  \end{tabular}
\end{lemma}

Now our
 sufficient condition for occur-check freeness is:

\begin{corollary}
\label{cor.tidy}
\label{corollary.tidy}
  A tidy program with a tidy query is occur-check free, under any selection
  rule. 
\end{corollary}

\noindent
{\sc Proof }
By Lemma \ref{lemma.tidy}, 
in each SLD-derivation for a tidy program and query, 
each query is tidy.
Assume $A$ is an atom from a tidy query and $H$ is the head of a standardized
apart tidy clause.  As $A$ is input-output disjoint and output linear and $H$
is input linear, 
$\{A{\doteq}H\}$ is NSTO, by Lemma \ref{lemma.nsto}.
\hfill $\Box$.

\subsection{Examples}
\label{sec.tidy.examples}

\noindent
Apt and Pellegrini \cite{AptP94-occur-check} found that 
the approaches based on well-modedness or nice modedness are inapplicable
to some programs, and introduced a more sophisticated approach.
The programs are 
{\sc flatten} \cite[Program 15.2]{Sterling-Shapiro-shorter},
{\sc quicksort\myunderscore d{}l} \cite[Program 15.4]{Sterling-Shapiro-shorter},
and {\sc normalize} (\cite[Program 15.7]{Sterling-Shapiro-shorter});
they employ difference lists.
Here we focus on {\sc flatten}, which flattens a given list.
(We split arguments of the form $t\mybackslash u$ or $t\mbox{\small++}u$ 
into two argument positions; the argument positions of
 built-in predicates are considered input.)
\pagebreak[3]
\[
\quad\ 
\begin{array}[t]{l}
\makebox[0pt]{\sc flatten:}
\\
    \%\,
    {\it flatten\_d l(X s,Y s,Zs) \mbox{ -- difference list }Y s\mybackslash Zs
    \mbox{ represents the flattened list }X s}
\\[.5ex]
{ \it
      flatten\_d l([X|X s],Y s, Zs) \gets
        flatten\_d l(X,Y s, Y s1),\
        flatten\_d l(X s,Y s1, Zs).
    }
    \\{ \it
      flatten\_d l(X,[X|X s], X s) \gets
         constant(X),\ X \mathrel{\mybackslash\mbox{\tt==}} [\,].
    }
    \\{ \it
      flatten\_d l([\,],X s, X s).
    }
\\[.5ex]
    {\it
      flatten(X s,Y s) \gets flatten\_d l(X s,Y s,[\,]).
    }
\end{array}
\]

\noindent
We see that, 
for the program to be tidy,
 the second and the third arguments of ${\it flatten\_d l}$ cannot be
both output (${\it Y s1}$ occurs in these positions in a clause body, which must
be output linear).  Also, the first and the second arguments of 
${\it flatten\_d l}$ cannot be both input, the same for the second and the
third one (as a clause head must be input linear).
The reader is encouraged to check that {\sc flatten} is tidy
under moding 
$M_1 = {\it flatten(+,-),\, flatten\_d l(+,-,+)}$,
and under
$M_2 = {\it flatten(-,+),\, flatten\_d l(-,+,-)}$.
The relation $\to_Q$ 
(where $Q=A_1,A_2$ is the body of the first clause)
consists of one pair; for $M_1$ it is $A_2\to_Q A_1$, 
for $M_2$ it is $A_1\to_Q A_2$.
In both cases the program remains tidy if the moding of {\it flatten} is
replaced by $(-,-)$. 

For any term $t$ and variable $R\not\in\Var(t)$, 
query $Q_0={\it flatten(t,R)}$ is tidy under $M_1$.
  (To be tidy under $M_2$, $Q_0$ has to be linear.)
By Corollary \ref{cor.tidy}, {\sc flatten} with $Q_0$ is occur-check free,
under any selection rule.

We only mention that  {\sc quicksort\myunderscore d{}l}
and {\sc normalize} are also tidy, and thus are occur-check free for a wide
class of queries.  
{\sc normalize} is similar to  {\sc flatten}, and is tidy for similar modings.
{\sc quicksort\myunderscore d{}l} is tidy 
for instance for modings\,
${\it quick sort(+,-)}$,\,\,${\it quick sort\_d l(+,-,+)}$, ${\it partition(+,+,-,-)}$,
\,and\,
${\it quick sort(-,+)}$, ${\it quick sort\_d l(-,+,-)}$, ${\it partition(-,+,+,+)}$.
Another example of a tidy program is {\sc derivative} (Example
\ref{ex.derivative} below).
 {\sloppy\par}

Surprisingly, the approach based on nice modedness {\em is} applicable to
{\sc flatten} and {\sc normalize}.  
{\sc flatten} is nicely moded under $M_2$, so is the query $Q_0$, provided it is
  linear.  As the clause heads are input linear, it follows that {\sc flatten}
  with $Q_0$ is occur-check free for the Prolog selection rule
  (by \cite[Corollary 7.25]{apt-prolog}).
Similarly,  {\sc normalize} is nicely moded
  (e.g.\ under ${\it normalize\_d s(-,+,-)}$);
we skip further details.
In both cases the modes may be seen as not natural; what is understood as
input data appears in a position moded as output.
This may explain why the nice modedness of the programs was not noticed 
in \cite{AptP94-occur-check}.

\section{Weakening NSTO}
The discussion on avoiding the occur-check above, and in all the
 work referred here, is based on the notion of NSTO.
We show that NSTO is a too strong requirement.  
Unification without the occur-check produces correct results also for some
pairs of atoms which are not NSTO.
In such cases the algorithm may temporarily construct infinite terms, but 
eventually halt with failure.  An example of such pair is
$p(a, f(X), X )$, $p(b, Y, Y )$.  For this pair, some runs of MMA
 halt due to 
selecting $a\mathop{\doteq}b$, some other ones due to a successful occur-check.
Omitting the occur-check would result in failure on $a\mathop{\doteq}b$;  
this is a correct result.

NSTO requires that each run of MMA does not perform action (6).  In this
section we show that it is sufficient that there {\em exists} such run.
For this we need to introduce a precise description of the algorithm without
the occur-check, called MMA$^-$.
We define WNSTO, a weaker version of NSTO, and show that MMA$^-$ produces
correct results for expression pairs that are WNSTO.
Then we show an example of a program with a query which is not occur-check
free, but 
will be correctly executed without the occur-check, as all the atom pairs to
be unified are WNSTO.
Then we present
sufficient conditions, based on WNSTO, for safely skipping the occur-check.

\subsection{An algorithm without the occur-check}
\noindent
By abuse of terminology, we will write
``unification algorithm without the occur-check'', despite such algorithm
does not correctly implement unification.
We would not consider any actual unification algorithm of Prolog,
this would require to deal with too many low level details.
See for instance the algorithm of
\cite[Section\,2]{DBLP:books/mit/AitKaci91}.
Instead, we use a more abstract algorithm, obtained from MMA.
We cannot simply drop the occur-check from MMA (by removing action (6) and 
the condition $X\not\in\Var(t)$ in action (5)).  The resulted algorithm 
may not terminate, as an equation $X\mathop{\doteq} t$ where $X\in\Var(t)$ can be selected
infinitely many times.  

We obtain a reasonable algorithm in two steps. 
First, MMA is made closer to
actual unification algorithms.  
The idea is to abandon action (5), except for $t$ being a variable.
(The action applies $\{X/t\}$ to all equations except one).
Instead, 
$\{X/t\}$ is applied only when needed, and only to one occurrence of $X$.
This happens when the variable becomes the left hand side
of two equations $X\myeq t,\ X\myeq u$
(where $t,u$ are not variables, and $X\not\in\Var(t,u)$).
Then $X\myeq u$ is replaced by $t\myeq u$.
To achieve termination, $t$ should not be larger than $u$.
From such algorithm the occur-check may be dropped.

Without loss of generality we assume that we deal with unification of terms.
Let $|t|$ be the number of occurrences in $t$ of variables and
function symbols, including constants.

\pagebreak[3]
\begin{definition}[{\rm\cite{Colmerauer1982}}]
{\bf MMA$\!^{\boldsymbol-}$} (MMA without the occur-check) is obtained from 
MMA by
\[
\begin{tabular}{l@{ }l}
(a) & removing action (6), and 
\\
(b) & replacing action (5) by
\\[1ex]
\multicolumn{2}{l}
{%
  \begin{tabular}{l@{ }l}
    (5a) &
    \parbox[t]{.75\textwidth}{
    $X\mathop{\doteq}Y$, where $X,Y$ are distinct variables and $X$ occurs elsewhere
\\
      \mbox{}\qquad$\longrightarrow$ \qquad 
      \begin{tabular}[t]{l}
      apply substitution $\{X/Y\}$ to all other equations, \\
      \end{tabular}
    }
  \end{tabular}%
}
\\[2.5ex]
\multicolumn{2}{l}
{%
  \begin{tabular}{l@{ }l}
    (5b) &
    \parbox[t]{.85\textwidth}
{
      $X\mathop{\doteq}t$, $X\mathop{\doteq}u$ 
where $t,u$ are distinct non-variable terms;
    let $\{s_1,s_2\}=\{t,u\}$ and $|s_1|\leq|s_2|$
       \\
       \mbox{}\qquad$\longrightarrow$ \qquad
      \begin{tabular}[t]{l}
      replace  $X\mathop{\doteq}s_2$  by $s_1\mathop{\doteq}s_2$
      \end{tabular}
    }
  \end{tabular}%
}
\end{tabular}
\]
A set $E$ of equations is in a {\bf semi-solved form} if $E$ is
$\{X_1\mathop{\doteq}t_1,\ldots,X_n\mathop{\doteq}t_n\}$, where $X_1,\ldots,X_n$ are distinct,
each $X_i$ is distinct from $t_i$,
and if a $t_i$ is a variable then $X_i$ occurs only once in $E$
(for $i=1,...,n$).
 $E$ is in a  {\bf solved form} 
 if, additionally, $X_i\not\in\Var(t_j)$ for $1\leq i,j\leq n$.
\end{definition}

\noindent
Note that inability of performing any action of MMA$^-$ means that the
equation set is in a semi-solved form.

Let us first discuss termination of MMA$^-$.
Let $k>1$ be an integer greater than the arity
of each function symbol appearing in the equation set $E$
which is the input of the algorithm. 
We define a function $||\ ||$ assigning natural numbers to
equation sets and equations:
\[
\begin{array}{c}
  || \{\, t_1{\doteq}u_1,\ldots, t_n{\doteq}u_n \,\} || = 
  \displaystyle\sum_{i=1}^n ||t_i{\doteq}u_i||,  \qquad \mbox{where}
\qquad
  ||t{\doteq}u|| = k^{\mbox{\,\small$\max(|t|,|u|)$}}.
\end{array}
\]
Assume that an action of MMA$^-$ is applied to a set $E$ of equations,
resulting in $E'$.
Then $||E||=||E'||$ if the action is (4), (5a), or (5b), and 
$||E||>||E'||$ if it is (3).
By the lemma below, $||E||>||E'||$ for action (1).
\begin{lemma}
\label{lemma.norm}
For any terms (or atoms) $s=f(\seq s),\ t=f(\seq t)$,
\ \ 
$
 ||  s{\doteq} t||>\displaystyle \sum_{i=1}^n ||s_i{\doteq}t_i||.  
$

\end{lemma}
{\sc Proof%
\footnote{%
  Function $||\, ||$ is proposed and
  the lemma is stated without proof in \cite{Colmerauer1982}.
  There, however, $k$ is
  the maximal arity of symbols from $E$, which does not
    make sense when it is 0 or 1. 
    The lemma also holds (with a slightly longer proof) for $k$ being the
    maximal number out of 2 and the arities of the symbols.
} %
 } %
The inequality obviously holds for $n=0$.  Let $n>0$, and
$l,r$ be, respectively, the left and the right hand side of the inequality.
Without loss of generality, assume that $|s|\geq |t|$.
Now
$l = k\cdot k^{|s_1|}\cdots k^{|s_n|} \geq k\cdot k^{|t_1|}\cdots k^{|t_n|}$.
Hence
$l/k\geq k^{|u|}$ for any $u\in\{\seq s, \seq t \}$.
Thus
$l/k\geq ||s_i{\doteq}t_i ||$ for $i=1,\ldots,n$, and then
$n\cdot l/k\geq r$.  As $n/k<1$, we obtain $l>r$.\,%
\hfill
$\Box$

 Let $f_{45a}(E)$ be  the number of those equations from $E$
to which action (4) or (5a) applies.
Let $f_{5b}(E)$ be  the number of equations of the form $X{\doteq}u$ in $E$,
where $u$ is not a variable.
Note that applying (4) or (5a) decreases $f_{45a}(E)$, 
and applying (5b) decreases $f_{5b}(E)$ without changing 
$f_{45a}(E)$, 

Now consider the lexicographic ordering $\prec_3$ on $\NN^3$ (cf. for
instance \cite[p.\,33]{apt-prolog}).
If $E'$ is obtained from $E$ by applying one action of the algorithm,  
it holds that
\[
 (||E'||, f_{45a}(E'), f_{5b}(E')) \prec_3  (||E||, f_{45a}(E), f_{5b}(E)).
\]
Thus, as $\prec_3$ is well-founded, 
MMA$^-$ terminates for any input set of equations $E$.

In discussing further properties of the algorithm,
we will consider possibly infinite terms ({\bf i-terms}) over the given
alphabet. 
We require that 
the set of variables occurring in an i-term is finite. 
The corresponding generalization of the
notion of substitution is called {\bf i-substitution}.

\begin{definition}
\label{def.i-}
A substitution (respectively i-substitution) $\theta$ is a {\bf solution}
({\bf i-solution}) of an equation $t\mathop{\doteq}u$ if
$t\theta=u\theta$;
\,\,$\theta$ is a {\em solution} ({\em i-solution}) of a set $E$ of equations, 
if $\theta$ is a {solution} ({i-solution}) of each equation from $E$.
Two sets of equations are {\bf equivalent} (respectively {\bf i-equivalent})
  if they have the same set of solutions (i-solutions).
\end{definition}

\begin{lemma}
\label{lemma.equivalent}  
Each action of MMA or of MMA$^-$ replaces an equation set by an i-equivalent
one.

\end{lemma}
{\sc Proof }
For any i-substitution $\theta$,  $f(\seq s)\theta= f (\seq t)\theta$ \,iff\,
$s_i=t_i$ for all $i\in\{1,\ldots,n\}$.  Thus the claim holds for action (1).
For actions (3), (4) the claim is obvious; the same for (2), (6),
Actions (5) and (5a) replace an equation set
$E=E_X\cup E_1$ by $E'=E_X\cup E_1\{X/t\}$, where $E_X=\{X\mathop{\doteq}t\}$.
Consider an i-solution $\theta$ of $E_X$.
So $X\theta = t\theta$.
Hence $(V\{X/t\})\theta = V\theta$ for any variable $V$,
and thus $t\{X/t\}\theta=t\theta$ for any expression $t$.
So $\theta$ is a solution of $E_1$ iff  $\theta$ is a solution of 
$E_1\{X/t\}$.
For (5b), 
equivalence of $\{X{\doteq}t,X{\doteq}u\}\cup E_1$ and $\{X{\doteq}t,t{\doteq}u\}\cup E_1$ 
follows immediately from Def.\ \ref{def.i-}
\hfill$\Box$

\begin{lemma}
\label{lemma.semi.solved}  
Any set of equations $E$ in a semi-solved form has an i-solution.
\end{lemma}
{\sc Proof }
If an equation of the form $X_i{\doteq}Y$ occurs in $E$ then
$E$ has an i-solution iff $E\setminus\{X_j{\doteq}Y\}$ has an i-solution
(as such $X_j$ occurs in $E$ only once), 
Hence we can assume that $E$ does not contain any equation of this form.
Now the result follows from Th.\,4.3.1 of 
\cite{DBLP:journals/tcs/Courcelle83}.
\hfill $\Box$
\medskip

It remains to discuss the results of MMA$^-$.  
Note that if $E$ and $E'$ are i-equivalent then they are equivalent$\!$. %
Consider a run $R$ of  MMA$^-$ starting from an equation set $E$.
If $R$ halts with failure (due to action (2)) then, by
Lemma \ref{lemma.equivalent}, $E$ has no solutions (is not unifiable).
If it halts with equation set $E'$ in semi-solved form, then
by Lemma \ref{lemma.equivalent}, $E$ is unifiable iff $E'$ is.
So applying MMA to $E'$, which boils down to applying actions (5) and (6),
either halts with failure, or produces a solved form $E''$, representing
an mgu of $E$.  
Prolog does not perform the occur-check, and treats the semi-solved form
as the result of unification.  
Prolog implementations present the result to the user in various ways.
For instance the answer to query
$g(X,X)=g(Y,f(Y))$ is displayed as $X {=} Y,\, Y {=} f(Y)$ by SWI, and as
$X{=} f(f(f(\ldots))),\, Y{=} f(f(f(\ldots)))$ by SICStus
(predicate =/2 is defined by clause ${=}(Z,Z)$).$\!\!$%

\subsection{WNSTO}

\noindent
Let us say that a run of MMA is {\em occur-check free} if the run does not
perform action (6). (In other words, 
no equation $X=t$ is selected where $X\in\Var(t)$ and $X\neq t$;
simply -- the occur-check does not succeed in the run).
An equation set $E$ is {\bf WNSTO} (weakly NSTO) when there exists an
occur-check free run of MMA for $E$.
When $E$ is $s\myeq t$ we also say that the unification of $s$ and $t$ is WNSTO.
A program $P$ with a query $Q$ is {\bf weakly occur-check free} if, under a
given selection rule, 
all the unification{}s available in the SLD-tree for $P$ with $Q$ are WNSTO.
A run of MMA$^-$ on an equation set
 $E$ is {\bf correct} if it produces correct results
i.e.\ the run halts with failure if $E$ is not unifiable, and produces a
unifiable equation set $E'$ in a semi-solved form otherwise.
The latter means that applying to $E'$ action (5) iteratively produces an mgu
of $E$, in a form of an equation set in a solved form.
We say that  MMA$^-$ is {\bf sound} for $E$
if all the runs of  MMA$^-$ on $E$ are correct.

Now we show that if unification of $E$ can be split in two parts and 
each of them is NSTO, then $E$ is WNSTO.
(For a proof see Appendix \ref{app.proof.lemma.iteration}.)

\begin{lemma}
  \label{lemma.iteration}
  Let $E_1\cup E_2$ be an equation set.

  If $E_1$ is not unifiable and is NSTO then  $E_1\cup E_2$ is WNSTO.

  If  $\theta_1$ is an mgu of $E_1$, and each
 $E_1$ and $E_2\theta_1$ is NSTO then  $E_1\cup E_2$ is WNSTO.

\end{lemma}

\begin{corollary}
\label{corollary.iteration}
    Consider a moding and atoms $p(s;t)$ and $p(s';t')$, where 
    $s\MYEQ s'$ is NSTO.

    If  $s\MYEQ s'$ is not unifiable then 
    $p(s;t)\myeq p(s';t')$ is WNSTO.

    If $\theta$ is an mgu of $s\MYEQ s'$, and
    $(t\MYEQ t')\theta$ is NSTO
    then  $p(s;t)\myeq p(s';t')$ is WNSTO.

\end{corollary}
{\sc Proof }
Equation $p(s;t)\myeq p(s';t')$ is WNSTO iff equation set 
$s\MYEQ s'\cup t\MYEQ t'$ is WNSTO.
Now Lemma \ref{lemma.iteration} applies.
\hfill$\Box$

\medskip
WNSTO is sufficient for the unification without the occur-check to work
correctly:
\begin{theorem}
\label{th.MMAminus}
  Consider an equation set $E$.  Assume that there exists an occur-check free
run of MMA on $E$.
Then  MMA$^-$ is sound for $E$. 
\end{theorem}
{\sc Proof }
Let $R_1$ be an occur-check free run of MMA on $E$, and $R_2$ be a run
of MMA$^-$ on $E$.  
We show that $R_2$ is correct.
Let $S$ be the set of the i-solutions of $E$, and thus of every equation set
$E'$ appearing in $R_1$ or $R_2$ (by Lemma \ref{lemma.equivalent}).

If $R_1$ succeeds then $S$ contains unifiers of $E$, and of every $E'$
appearing in $R_2$.  Hence action (2) is not performed in $R_2$,
and $R_2$ halts with success producing a unifiable equation set $E_2$ in a
semi-solved form.

If $R_1$ halts with failure then the last performed action is (2), 
thus $S=\emptyset$.
This implies that $R_2$ does not produce a semi-solved form
(by Lemma \ref{lemma.semi.solved}).  Hence $R_2$ terminates with failure,
due to action (2).
\hfill$\Box$

\medskip
It immediately follows that a weakly occur-check free program can be safely
executed without the occur check:

\begin{corollary}
\label{corollary.main}
Assume a selection rule.
If a program $P$ with a query $Q$ is weakly occur-check free
then algorithm MMA$^-$ is sound
for each unification available in the SLD-tree for $P$ with $Q$.
\end{corollary}

In other words, $P$ with $Q$ may be correctly executed without the occur-check.

\subsection{Example -- a weakly occur-check free program}
\label{sec.nqueens}
The core fragment of the $n$ queens program \cite{Fruehwirth91} will 
be now used as an example. We call it \nqueens, see 
\cite{drabent.nqueens.tplp.pre}
 for explanations.

\vspace{\abovedisplayskip}
\noindent
\mbox{}\hfill%
\begin{minipage}[t]{.8\textwidth}
{\small %
\begin{verbatim}
    pqs(0,_,_,_).
    pqs(s(I),Cs,Us,[_|Ds]):-
            pqs(I,Cs,[_|Us],Ds),
            pq(s(I),Cs,Us,Ds).

    pq(I,[I|_],[I|_],[I|_]).
    pq(I,[_|Cs],[_|Us],[_|Ds]):-
            pq(I,Cs,Us,Ds).
\end{verbatim}
}
\end{minipage}%
\hfill
 \begin{minipage}[t]{.035\textwidth}
\raggedleft
\small%
(\refstepcounter{equation}\theequation\label{clause1})
\\ \ \\
       \refstepcounter{equation}%
      (\theequation\label{clause2})%
\\ \ \\ \ \\ 
       \refstepcounter{equation}%
      (\theequation\label{clause3})
\\[1.5ex]
       \refstepcounter{equation}%
        {(\theequation\label{clause4})}
\end{minipage}%
\vspace{\belowdisplayskip}
\\
A typical initial query is 
$Q_{\rm in}={\it p q s}(n,q_0,\myunderscore,\myunderscore)$, where
$q_0$ is a list of distinct variables, and $n$ a natural number represented
as $s^i(0)$.
The program works on non-ground data.

We now show that the standard syntactic approaches to deal with avoiding the
occur-check are inapplicable to \nqueens.
Under no moding the program is
well-moded with $Q_{\rm in}$ because its answers are non-ground.
To be tidy (or nicely moded with input linear clause heads), 
at most one position of ${\it p q}$ is input (as 
(\ref{clause3}) must be input linear). 
Thus at least three positions of ${\it p q s}$ have to be output
(as a variable from an input position of the head
of (\ref{clause2}) cannot appear in an output position of body atom
   ${\it p q( s(I), Cs, Us, D s)}$).  This makes the body not output linear,
contradiction.

It can be shown that \nqueens with $Q_{\rm in}$ is occur-check free under any
selection rule, by showing that in all SLD-derivations each atom in each query
is linear \cite{drabent.occur-check.report}. 
This is however rather tedious.
The program is not occur-check free for some non linear queries, for instance
for $A_{\rm S TO}={\it p q}(m, L, [L|\myunderscore],\myunderscore)$ (where $m$ is ground).
This is because unifying $A_{\rm S TO}$ with the unit clause  (\ref{clause3})
is not NSTO. 

We now show that \nqueens can be correctly executed without the occur-check,
for a wider class of initial queries, including each 
${\it p q s}(m, t_1, t_2, t_3)$ where $m$ is ground.
Let us say that a query $Q$ is {\em 1-ground} 
\phantomsection
\label{place.1-ground}
if the first argument of the predicate symbol in each atom of $Q$ is ground.
We show that:

\begin{proposition}
 \nqueens is weakly occur-check free, under any selection
rule, for any 1-ground query.  
\end{proposition}

\noindent
 {\sc Proof }
Note first that each query in each SLD-derivation is 1-ground.
Let $A=p(\seq[4]s)$ be a 1-ground atom, and
 $H={\it p q}(I, [I|\myunderscore], [I|\myunderscore], [I|\myunderscore])$
be the head of (\ref{clause3}), standardized apart.
Equation $s_1\myeq I$
  is NSTO and $\theta=\{I/s_1\}$ is its mgu.
Let $s=(s_2,s_3,s_4)$ and 
$t=([I|\myunderscore], [I|\myunderscore], [I|\myunderscore])$.
As $s_1$ is ground, $t\theta$ is linear.
Thus $s\theta\MYEQ t\theta$ is NSTO by Lemma \ref{lemma.nsto}.
Hence by Lemma  \ref{lemma.iteration}, $s_1,s\mathop{\MYEQ} I,t$ is WNSTO.
So $A\myeq H$ is WNSTO.
The cases of the remaining clause heads of \nqueens are obvious, as the
heads are linear.
For another proof, see Examples \ref{ex.nqueens.free},
\ref{ex.nqueens.free.syntactic}.
\hfill $\Box$
{\sloppy\par}

\medskip

By Corollary \ref{corollary.main},
\nqueens with with any 1-ground query is correctly executed
without the occur-check, under any selection rule.

\nqueens may be considered a somehow unusual program.  However
similar issues appear with rather typical programs dealing with ground data.
Assume, for instance, that data items from a ground data structure are to be
copied into two data structures.  Program
\[
\phantomsection
\label{place.use2}
\mbox{\sc use2:}
\quad
{\it
p( [X|X s], \, f(X,X s1),\,  [g(X,\myunderscore)|X s2]
 ) \gets p(X s,X s1,X s2).
\qquad
p([\,],g,[\,]).
}
\]
provides a concise example.
Similarly as for \nqueens, it can be shown that
{\sc use2} is not occur-check free for some 1-ground
queries,  but is weakly occur-check free for all such queries.%
\footnote{{\sc use2}
    is well-moded under $p(+,-,-)$, but the approach for well-moded programs
    does not apply, as the clause head is not output linear.
 In contrast to \nqueens, {\sc use2} can be treated as tidy, or nicely moded.
 The program is tidy under any moding with at most one position $+$.
 Hence it is occur-check free for tidy queries
 (they are a proper subset of 1-ground queries, and include all linear
 queries). 
} 
For a quick proof see Ex.\,\ref{ex.nqueens.free} or 
\ref{ex.nqueens.free.syntactic}.

\subsection{Sufficient conditions for WNSTO}
\label{sec.sufficient.wnsto}
Now we discuss sufficient conditions for safely avoiding the occur-check
due to WNSTO.
We assume that the programs dealt with are moded.

We say that a selection rule is {\bf compatible with moding}
(for a program $P$ with a query $Q$)
if (i) the input positions are ground in each selected atom
in the SLD-tree for $P$ with $Q$
(\cite{AptL95.delays} calls this ``delay declarations imply the moding''), and
(ii) some atom is selected in a query whenever the query contains an 
atom with its input positions ground.
Note that (i) implies that the selection rule is partial,
in the sense that there exist nonempty queries in which no atom is selected.
For such a query no resolvent exists,
this is called {\em floundering} (or {deadlock}).

An atom $A$ is {\bf weakly linear} if any variable $X$ which occurs more than
once in $A$ occurs in an input position of $A$.
(Speaking informally, grounding the variables in the input positions of $A$
results in a linear atom.)

\begin{lemma}
\label{lemma.wnsto}
Consider variable disjoint atoms $A$ and $H$, such that the input positions
of $A$ are ground, and $H$ is weakly linear.  The unification of $A$ and $H$
is WNSTO.
\end{lemma}
{\sc Proof }
Let $A=p(s;t)$, where $s$ is ground, and $H=p(s';t')$.  Equation set
$s\MYEQ s'$ is NSTO (by Lemma \ref{lemma.nsto}).
Assume that $s\MYEQ s'$ is unifiable and that $\theta$ is an mgu of $s\MYEQ s'$.
Thus $X\theta$ is ground for each variable $X\in\Var(s')$.
Hence $t'\theta$ is linear, and $(t\MYEQ t')\theta$ is NSTO 
(by Lemma \ref{lemma.nsto}). 
Now by Corollary \ref{corollary.iteration},
$A\myeq H$ is WNSTO.
   \hfill$\Box$

\medskip
It immediately follows:
\begin{corollary}
\label{corollary.weakly..free}
  Let $P$ be a program in which each clause head is weakly linear.
  If the selection rule is compatible with moding then $P$ (with any query)
  is weakly occur-check free.  
\end{corollary}

\vspace{-1\medskipamount}
\begin{example}\rm
\label{ex.nqueens.free}

The heads of the clauses of 
\nqueens are weakly linear 
under moding ${\it p q s}(+,-,-,-)$, ${\it p q}(+,-,-,-)$.
By Corollary \ref{corollary.weakly..free}, the program
(with any query) is weakly occur-check free under
any selection rule compatible with moding.
Consider a query $Q$ which is 1-ground
(cf.\ Section \ref{sec.nqueens}, p.\,\pageref{place.1-ground}).
A simple check shows that in any SLD-derivation for \nqueens and $Q$ all
queries are 1-ground.  So 
each selection rule is compatible with moding (for \nqueens with $Q$).
Thus \nqueens with any 1-ground query is weakly occur-check free
for any selection rule.
 \ %
The same reasoning applies to {\sc use2}, with $p(+,-,-)$.

\end{example}

Now we provide a syntactic sufficient condition for a program to
be weakly occur-check free.
It employs a generalized notion of moding, in which some argument positions 
may be neither $+$ (input) nor $-$ (output),
to such positions we assign $\bot$ (neutral).
We will call it {\em\bf 3-moding} when it is necessary to distinguish it from
a standard moding.
We write $p(s;t;u)$ to represent an atom $p(\seq t)$ and to state that
$s$ (respectively $t$, $u$) is the sequence of terms in its $+$ ($-$,
$\bot$) positions.
The idea is to distinguish (as $+$ or $-$) some argument positions which,
roughly speaking, deal with ground data.
A syntactic sufficient condition will
imply for LD-derivations that in each selected atom the input positions are
ground.

By a {\bf well-3-moded} program (or query) we mean one which 
becomes well-moded after removing the $\bot$ argument positions.
For a direct definition, let a {\em defining occurrence} of a variable $V$ in
a clause $C = H{\gets} Q$ be an occurrence of $V$ in an input position of
H, or in an output position in $Q$.
Now $C$ is {\em well-3-moded} 
when each variable $V$ in an output position of $H$ has its
defining occurrence in $C$, and each occurrence of a $V$ in an
input position in $Q$ is preceded by a defining occurrence of $V$
in another literal of $C$
\cite{Dra87}.
A query $Q$ is well-3-moded when clause $p\gets Q$ is.
An equivalent definition can be obtained by an obvious adaptation of 
\cite[Def.\,7.8]{apt-prolog}.
Note that any query with its input positions ground is well-3-moded.

We now use the fact that
well-3-moded programs/queries inherit the main properties of well-moded ones.

\begin{lemma}
\label{lemma.well-3-moded}
  Let $P$ and $Q$ be well-3-moded. %

1.\mbox{ }%
All queries in SLD-derivations of $P$ with $Q$ are well-3-moded. %

2.\mbox{ }%
\begin{parbox}[t]{.9\textwidth}{
    For $P$ with $Q$
    the Prolog selection rule is compatible with moding.
}
\end{parbox}
3.\mbox{ }%
\parbox[t]{.938\textwidth}{
If each clause head in $P$ is weakly linear then
$P$ with $Q$ is weakly occur-check free under the Prolog selection rule 
(and any selection rule compatible with moding).
Moreover, if no argument position is moded as output then 
$P$ with $Q$ is weakly occur-check free under any selection rule.
}%
\vspace{+.9ex}

4.\mbox{ }%
$P$ with $Q$ does not flounder under any selection rule compatible with moding.

\end{lemma}

\noindent
{\sc Proof} %
\ 1.\,An SLD-resolvent of a well-3-moded query and a well-3-moded clause 
is well-3-moded. The proof is the same as that of the analogical property 
of well-moded queries and clauses \cite[Lemma 7.9]{apt-prolog}.
\ 2.\,and 4.\,\,From 1.\,and the fact that the
 input positions of the first atom of a  well-3-moded query are ground.
\ 3.\,From 2.\,by Corollary \ref{corollary.weakly..free}.
Additionally, 
 under a 3-moding without $-$, all input positions in a well-3-moded query are
 ground.  Thus each selection rule is compatible with moding and 
 Corollary \ref{corollary.weakly..free} applies.
\hfill $\Box$

\begin{example}\rm
\label{ex.nqueens.free.syntactic}
Programs
\nqueens and {\sc use2} are well-3-moded under 
${\it p q s}(+,\bot,\bot,\bot)$, ${\it p q}(+,\bot,\bot,\bot)$, and
$p(+,\bot,\bot)$; so is any 1-ground query.  Their clause heads are weakly
linear. 
(The same holds under the modings from Ex.\,\ref{ex.nqueens.free}.)
Thus by Lemma \ref{lemma.well-3-moded}.3, the programs are 
weakly occur-check free for 1-ground
queries, under any selection rule.
So we obtained by syntactic means
the results of Ex.\,\ref{ex.nqueens.free}.

\end{example}

\begin{example}\rm
\label{ex.derivative}

\noindent
Apt and Pellegrini \cite{AptP94-occur-check} use program {\sc derivative} 
\cite[Program 3.30]{Sterling-Shapiro-shorter}
as an example for an approach combining those for well-moded and nicely moded
programs.  
Here are representative clauses of the program
(infix operators ${\uparrow},{*}$ are used).
\[
\mbox{\sc derivative}\colon \qquad
\begin{array}[t]{l}
  d(X,X,s(0)). \\
  d(X{\uparrow} s(N),\, X,\, s(N){*}X{\uparrow} N ).
\\
    {\it d( F{*}G,\ X,\, F{*}D G {+} D F{*}G ) \gets  d(F,X,D F), d(G,X,D G).
      }
\end{array}
\]
\noindent
A typical query is $d(e,x,t)$, where $e,x$ are ground ($e$ represents
an expression, and $x$ a variable), $t$ is often a variable.
Here moding $d(+,\bot,\bot)$ will be sufficient.
Consider a query $Q=d(e_1,x_1,t_1),\ldots, d(e_n,x_n,t_n)$
where $\seq e$ are ground.  
{\sc derivative} and $Q$ are well-3-moded moded under $d(+,\bot,\bot)$.
Also, the clause heads are weakly linear.
By Lemma \ref{lemma.well-3-moded}.3, 
the program with $Q$ is weakly occur-check free under any selection rule.
    Alternatively,
 {\sc derivative} is tidy under $d(-,+,-)$.
Hence, by Corollary \ref{corollary.tidy},
under any selection rule the program is occur-check free for tidy queries,
including any linear queries.

\cite{AptP94-occur-check} applied a combination of methods of well-moding and
nice moding to show that {\sc derivative} is occur-check free
for an atomic $Q$ ($n=1$) with ground $e_1,x_1$ and linear
$t_1$, under LD-resolution.
That result is subsumed by each of our two conclusions above.
Surprisingly, 
a more general result can be obtained by a simpler approach from that work.
Under $d(-,+,-)$ {\sc derivative} is nicely-moded
and its clause heads are input linear.  
Thus it is occur-check free under LD-resolution for any nicely moded queries,
this includes any linear queries.

\end{example}

In this section we dealt with clause heads whose certain instances are linear.
Appendix
\ref{app.weakly.tidy} employs clauses whose certain instances are tidy,
to construct another sufficient condition for weak occur-check freeness.

\section{Comments}

Let us first discuss briefly the limits of applicability of the presented results.
The approaches discussed here are based on conditions imposed
on clauses and queries.  The conditions treat any predicate argument as a
single entity, and refer to groundness or to placement of variables within 
certain argument positions.  
    %
%
This may not be sufficient when the occur-check depends on
other features of the terms in argument positions.
For instance, in the SAT-solver of Howe and King
\cite{howe.king.tcs-shorter} 
an argument is a non-linear list of lists of pairs, 
and for occur-check freeness the first element of each pair should be ground
\cite{Drabent.tplp18}.
In such case our methods fail, and some semantic analysis is needed 
instead.
One may expect that introducing a suitable type system could be useful.

Introduction of WNSTO has two consequences. 
Some cases where unification is not NSTO can actually be safely executed
without the occur check.
Also, reasoning based on WNSTO is sometimes simpler.  For instance,
showing that program \nqueens is occur-check free was substantially more
complicated than showing it to be weakly occur-check free for a
wider class of queries.

Most of the employed sufficient conditions are based on the notion of modes.
Examples show that modes (except for well-moded programs)
do not need to correspond to any intuitive 
understanding of data flow.  
Instead, they deal with how variables are placed in argument positions.
An output argument may well be used for input data. 
Neglecting this fact may be the reason why in some examples
of \cite{AptP94-occur-check} unnecessarily complicated methods were applied, 
or more general results could have been obtained.
(For examples and explanations see the comments
on  {\sc flatten} and  {\sc normalize} in Section \ref{sec.tidy.examples}, 
and on {\sc derivative} in Ex.\,\ref{ex.derivative}.)

\paragraph{Conclusions}
The main contribution of this paper is weakening the notion of NSTO (not
subject to occur-check) used in the previous work on avoiding the occur-check.
We generalize NSTO to WNSTO (weakly NSTO).
This leads to a generalization of the notion of occur-check free
programs/queries (based on NSTO) to {\em weakly occur-check free}
ones (based on WNSTO). 
We proved that unification without the occur-check is sound for any input which
is \mbox{WNSTO}. 
We presented a few sufficient conditions for WNSTO, and
for a program/query being weakly occur-check free.
Some conditions are syntactic, like   
Lemma \ref{lemma.well-3-moded},
some refer to semantic notions, like Corollary \ref{corollary.weakly..free}
which explicitly refers to details of SLD-derivations.
Additionally, we presented a sufficient condition based on NSTO, generalizing 
the approach based on nicely moded programs.
Examples show that the proposed approach makes it possible to omit the
occur-check in cases, to which the approaches based on NSTO are inapplicable.
In some other cases, it leads to simpler proofs.

\appendix
\DeclareRobustCommand{\lemmaiteration}{\ref{lemma.iteration}}
\section{Appendix. Proof of Lemma \lemmaiteration}
\label{app.proof.lemma.iteration}

The proof employs a technical lemma.
\begin{lemma}
  \label{lemma.MMA.mgu}
  Consider a run $R$ of MMA producing an mgu $\theta$.  Let 
  $X_1\myeq t_1,\ldots,X_k\myeq t_k$ (in this order) be the equations selected
  in action (5) in $R$, and $\gamma_i=\{X_1/ t_1\}$ for $i=1,\ldots,k$.  Then
   $\theta=\gamma_1\cdots\gamma_k$.  Moreover, the last equation set of $R$ is 
$\{X_i\myeq t_i\gamma_{i+1}\cdots\gamma_k \mid 0<i\leq k \,\}$.
\end{lemma}
{\sc Proof }
Action (5) means applying $\gamma_i$ to all the equations except for 
$X_i\myeq t_i$.  Moreover, (a current instance of) $X_i\myeq t_i$ is not
selected anymore in $R$.  So $\theta$ contains a pair 
$X_i/ t_i\gamma_{i+1}\cdots\gamma_k$.
Let $\psi_j=\{X_i/t_i\gamma_{i+1}\cdots\gamma_j \mid 0<i\leq j \,\}$ and
$\varphi_j=\gamma_{1}\cdots\gamma_j$, for $j=0,\ldots,k$.
Now $\psi_j=\varphi_j$, by induction on $j$ 
(as
$
\{X_i/t_i\gamma_{i+1}\cdots\gamma_j \mid 0<i\leq j \,\}\gamma_{j+1}=
\{X_i/t_i\gamma_{i+1}\cdots\gamma_{j+1} \mid 0<i\leq j \,\}\cup\gamma_{j+1}
$).
Thus $\theta=\varphi_k$.
\hfill$\Box$

\medskip\noindent
{\sc Proof }(of Lemma \ref{lemma.iteration}) \
Assume that $E_1$ is not unifiable.  Then from a run $R$ of MMA on  
$E_1$ one can construct in an obvious way a run $R'$ of MMA on 
$E_1\cup E_2$, performing the same actions and selecting the same equations.
Thus action (6) is not performed in $R'$.

Now assume that $\theta_1$ is an mgu of $E_1$.
Consider a run $R_1$ of MMA on $E_1$, and a run $R_2$ of MMA on $E_2\theta_1$.
Without loss of generality we may assume that $\theta_1$ is the result of
$R_1$.  (Otherwise, $R_1$ produces $\theta$ such that
$E_2\theta_1$ is a variant of $E_2\theta$.)

Let step (5) be applied in $R_1$ to equations 
$X_1\myeq t_1,\ldots,X_k\myeq t_k$ (in this order), 
and in  $R_2$ to 
$X_{k+1}\myeq t_{k+1},\ldots,X_m\myeq t_m$ (in this order).
Let $\gamma_i=\{X_i/t_i\}$, for $i=1,\ldots,m$.
Let $F$ be the last equation set in $R_1$;
by Lemma \ref{lemma.MMA.mgu},
$F=\{X_1\myeq t_1\gamma_2\cdots\gamma_k,\ldots,X_k\myeq t_k\}$.
{\sloppy\par}

To construct a run $R$ of MMA on $E_1\cup E_2$, 
for each equation set from $R_1$ or $R_2$ we construct a corresponding
equation set from $R$.

Consider an equation set $E$ in $R_1$.
Let $X_1\myeq t_1,\ldots,X_i\myeq t_i$ be the equations on which 
action (5) has been performed in $R_1$ until obtaining $E$
($i\in\{0,\ldots,k\}$). 
The equation set corresponding to $E$ is
$E' = E\cup E_2\gamma_1\cdots\gamma_i$.
In particular (by Lemma \ref{lemma.MMA.mgu}),
$F\cup E_2\theta_1$ corresponds to the last equation $F$ of $R_1$.

For $i\in\{k,\ldots,m\}$ let us define 
$F_i=\{X_l\mathrel{\myeq}t_l\gamma_{l+1}\cdots\gamma_i \mid 0<l\leq k \,\}$.
Note that $F_k=F$.
Consider an equation set $E$ in $R_2$.
Let $X_{k+1}\myeq t_{k+1},\ldots,X_i\myeq t_i$ be the equations on which 
action (5) has been performed in $R_2$ until obtaining $E$
($i\in\{k,\ldots,m\}$). 
The equation set corresponding to $E$ is
$E'=F_i\cup E$.

Consider now the sequence consisting of the equation sets corresponding to
those of
$R_1$ and then of the equation sets corresponding to those of $R_2$ (without
the first one, to avoid a repetition of $F\cup E_2\theta_1$).
The sequence is a run of MMA on $E_1\cup E_2$.
As the run does not involve action (6), $E_1\cup E_2$ is WNSTO.
\hfill$\Box$

\medskip

\section{Appendix. Another syntactic sufficient condition}
\label{app.weakly.tidy}
Here we present a sufficient condition for avoiding the
occur-check, related to WNSTO and 
based on the syntactic conditions for tidy programs.

Consider a 3-moding $M$ and an additional moding $M'$,
for the latter we use symbols $+',-'$.
Consider transforming each clause $C$ of a program $P$, by grounding the
variables 
that occur in the $+$ positions in the head.
Let $P'$ be the resulting program.
Let us say that $P$ is {\em weakly tidy} (under $M,M'$)
 if $P'$ is tidy under $M'$.

For an example, consider a program $P$ containing a clause $C$ with body
$B = q(X,Y),\linebreak[3] q(Y,Z),\linebreak[3] q(Z,X)$.
Assume also that $P$ contains a clause head $H=q(t,u)$ with 
$t,u$ containing a common variable.
For $P$ to be tidy, the argument positions of $q$ cannot be both input (due
to $H$), and cannot be both output (due to $B$).
However, if they are $(+,-)$, or $(-,+)$ then $\to_B$ is cyclic.
Thus $P$ is not tidy (and not nicely moded) under any moding.
Assume now that $C$ is $p(X)\gets B$.
Under $p(+)$, and $q(+',-')$
the clause is weakly tidy.  (A corresponding clause of $P'$ is
$ p(s) \gets q(s,Y),\linebreak[3] q(Y,Z),\linebreak[3] q(Z,s)$, for a
ground term $s$; note that $q(+',-')$ may be replaced by $q(-',+')$.)
By the lemma below,
if $p(s)$ is selected in a tidy query $Q$ then the resolvent of $Q$ and $C$
is tidy.

\begin{lemma}

Let $P$ be a weakly tidy program under $M,M'$, and $Q$ a query tidy under $M'$.
Under any selection rule compatible with $M$

each query in any SLD-derivation for $P$ with $Q$ is tidy under $M'$, and

$P$ with $Q$ is weakly occur-check free.

\end{lemma}

\noindent
{\sc Proof }
Let $A$ be the selected atom of a tidy (under $M'$) query $Q$, and $H$ be
the head of a standardized apart clause $C$ of $P$.
Let $A=p(s;t;u)$ and $H=p(s';t';u')$ (under $M$).
Unifying $A$ with $H$ can be divided in two steps:
  \cite[Lemma 2.24]{apt-prolog}.\linebreak[3] 

\  1.\ Unifying  $s$ with $s'$.
   As $s$ is ground, $s\MYEQ s'$ is NSTO.

\  2.\ Unifying $(t,u\MYEQ t',u')\theta$ 
provided that $s\MYEQ s'$ is unifiable with an mgu $\theta$.
This is the same as unifying $A\theta\myeq H\theta$
(as $s\theta=s'\theta$ and is ground).
Note that $Q\theta=Q$ (thus $A\theta=A$),
 and that $C\theta$ is tidy under $M'$.
So $A\theta$ is an atom from a tidy query, and $H\theta$ is the head of 
a standardized apart tidy clause $C\theta$. 
By Corollary \ref{corollary.tidy}, $A\theta\myeq H\theta$ is NSTO.

Now by Corollary \ref{corollary.iteration}, $A\myeq H$ is WNSTO.
If $A\myeq H$ is unifiable then, by Lemma \ref{lemma.tidy}, the resolvent
$Q'$ of 
$Q\theta$ and $C\theta$ is tidy.  $Q'$ is also the resolvent of $Q$ and $C$.

We showed, for a tidy query $Q$ and a standardized apart clause $C$ of $P$,
that unification of the selected atom of $Q$ with the head of $C$ is WNSTO,
and that the resolvent (if it exists) of $Q$ with $C$ is tidy.
The Lemma follows by simple induction. 
\hfill$\Box$

Consider $P,Q$ satisfying the conditions of the Lemma.
If $P$ and $Q$ are well-3-moded under $M$ then $P$ with $Q$ is weakly
occur-check free under Prolog selection rule (and under any selection rule
compatible with $M$, under such rule it does not flounder).

\bibliographystyle{eptcsalpha}
\bibliography{bibpearl,bibmagic,bibs-s,bibshorter,ja}

\end{document}